\documentclass[preprints,article,accept,moreauthors,pdftex]{Definitions/mdpi} 

\usepackage[T1]{fontenc}
\usepackage[utf8]{inputenc}
\usepackage{amsmath,amsthm,amssymb}
\usepackage{amssymb}
\usepackage{bm}
\usepackage{enumitem}
\usepackage{float}
\usepackage{circledsteps}

\firstpage{1} 
\pubvolume{1}
\issuenum{1}
\articlenumber{0}
\pubyear{2022}
\copyrightyear{2022}
\datereceived{} 
\dateaccepted{} 
\datepublished{} 
\hreflink{https://doi.org/} 

\Title{Constraining MOdified Gravity with the S2 star}

\TitleCitation{Constraining MOdified Gravity with with the S2 star}

\Author{Riccardo Della Monica $^{1,}$*\orcidA{}, Ivan de Martino $^{1}$\orcidB{} and Mariafelicia de Laurentis $^{2,3}$}

\AuthorNames{Riccardo Della Monica, Ivan de Martino and Mariafelicia de Laurentis}

\AuthorCitation{Della Monica, R.; de Martino, I.; de Laurentis, M.}

\address{%
$^{1}$Universidad de Salamanca, Departamento de Fisica Fundamental, P. de la Merced, Salamanca, ES\\
$^{2}$Dipartimento di Fisica, Universit\'a
di Napoli {}``Federico II'', Compl. Univ. di
Monte S. Angelo, Edificio G, Via Cinthia, I-80126, Napoli, Italy\\
$^{3}$ INFN Sezione  di Napoli, Compl. Univ. di
Monte S. Angelo, Edificio G, Via Cinthia, I-80126, Napoli, Italy
}

\corres{Correspondence: rdellamonica@usal.es}

\abstract{We have used publicly available kinematic data for the S2 star to constrain the parameter space of MOdified Gravity. Integrating geodesics and using a Markov Chain Monte Carlo algorithm we have provided with the first constraint on the scales of the Galactic Centre for the parameter $\alpha$ of the theory, which represents the fractional increment of the gravitational constant $G$ with respect to its Newtonian value.
Namely, $\alpha \lesssim 0.662$ at 99.7\% confidence level (where $\alpha = 0$ reduces the theory to General Relativity).}


\begin{document}


\section{Introduction}
Scalar-Tensor-Vector Gravity (STVG), also referred to in literature as MOdified Gravity (MOG), is a theory of gravity firstly proposed in \cite{moffat_svtg} as an alternative to Einstein's theory of General Relativity (GR). It introduces extra fields in the description of the gravitational interaction, allowing for correct predictions on galactic and extra galactic scales \cite{Brownstein2006, moffat_galaxyrot1, moffat_galaxyrot2, DeMartino2017, deMartino2020}, without resorting to dark matter \cite{deMartinoChakrabarty2020}. The gravitational action in MOG presents additional terms along the classical Hilbert-Einstein action, depending on the metric tensor $g_{\alpha\beta}$ of space-time. More specifically, a massive vector field $\varphi^\alpha$ is introduced and its mass, $\mu$, is treated as a scalar field. Furthermore, also Newton's gravitational constant $G_N$ is elevated to a scalar field $G$. 

The motion of test particles in MOG is affected by the presence of the vector field $\varphi^\alpha$ which acts as a fifth-force, whose repulsive character counteracts the increased attraction due to the scalar field nature of $G$. The fractional increment of $G$, with respect to its Newtonian value, $G_N$, is given by a new parameter of the theory, $\alpha = (G-G_N)/G_N$. A distinctive feature in the motion of test massive bodies in MOG is that Keplerian orbits in a central potential are characterized by an increased value of the rate of orbital precession \cite{DellaMonica2021_a,rdm}. This is given by:
\begin{equation}
    \Delta\omega_{\rm MOG} = \Delta\omega_{\rm GR}\left(1+\frac{5}{6}\alpha\right),
    \label{eq:stvg_precession}
\end{equation}
where, $\Delta\omega_{\rm GR}$ is the usual expression of the periastron advance in GR,
\begin{equation}
    \Delta\omega_{\rm GR} = \frac{6\pi G_NM}{ac^2(1-e^2)}\,,
    \label{eq:gr_precession}
\end{equation}
related to semi-major axis, $a$, and eccentricity, $e$, of the orbiting body.

Here, we will summarize the extended work done in \cite{DellaMonica2021_a}, where we used publicly available data for the S2 star from \cite{gillessen}, along with the measurement of its orbital precession from \cite{gravity} to constrain the parameter space of MOG.

\section{MOdified Gravity}

In MOG, the gravitational action is written as \cite{moffat_svtg}:
\begin{equation}
    \mathcal{S} = \mathcal{S}_{\rm HE}+\mathcal{S}_m+\mathcal{S}_V+\mathcal{S}_S.
    \label{eq:stvg-action}
\end{equation}
The first term, $\mathcal{S}_{\rm HE}$, is the classical Hilbert-Einstein action of GR, while $S_m$ is related to the ordinary matter energy-momentum tensor,
\begin{align}
    \mathcal{S}_{\rm HE} &= \frac{1}{16\pi}\int d^4x\sqrt{-g}\frac{1}{G}R,&
    T^m_{\alpha\beta} &= -\frac{2}{\sqrt{-g}}\frac{\delta\mathcal{S}_m}{\delta g^{\alpha\beta}}.
    \label{eq:hilbert-einstein}
\end{align}
where $g_{\alpha\beta}$ is the metric tensor of space-time, $g$ its determinant and $R$ the Ricci scalar. The two extra terms, $\mathcal{S}_V$ and $\mathcal{S}_S$, on the other hand, are related to the vector and scalar field respectively, and read:
\begin{align}
    \mathcal{S}_V =& -\int d^4x\sqrt{-g}\left(\frac{1}{4}B^{\alpha\beta}B_{\alpha\beta}-\frac{1}{2}\mu^2\varphi^\alpha\varphi_\alpha+V(\varphi)\right),\\
    \mathcal{S}_S =& \int d^4x\sqrt{-g}\frac{1}{G^3}\left(\frac{1}{2}g^{\alpha\beta}\nabla_\alpha G\nabla_\beta G-V(G)\right)+\int d^4x\frac{1}{\mu^2G}\left(\frac{1}{2}g^{\alpha\beta}\nabla_\alpha\mu\nabla_\beta\mu-V(\mu) \right).
\end{align}
With $\nabla_\alpha$ we have indicated the covariant derivative related to the metric tensor $g_{\alpha\beta}$, and with $B_{\alpha\beta}$ the Faraday tensor associated to the massive vector field $\varphi_\alpha$: $B_{\alpha\beta}=\nabla_\alpha\varphi_\beta-\nabla_\beta\varphi_\alpha \,$.
$V(\varphi)$, $V(G)$ and $V(\mu)$, on the other hand, represent scalar potentials describing the self-interaction of the vector and scalar fields.

In MOG, particles with mass $m$ move according to a modified version of the geodesic equations \cite{moffat_eqmotion}:
\begin{align}
    \left(\frac{d^2x^\alpha}{d\lambda^2}+\Gamma^\alpha_{\beta\rho}\frac{dx^\beta}{d\lambda}\frac{dx^\rho}{d\lambda} \right)=\frac{q}{m}{B^{\alpha}}_\beta\frac{dx^\beta}{d\lambda}.
    \label{eq:geodesic-equations}
\end{align}
The term on the right-hand side represents a fifth force \citep{moffat_svtg, moffat_galaxyrot1, moffat_eqmotion}, due to the coupling between massive particles and the vector field $\varphi^\alpha$. The coupling constant, $q$, is postulated to be positive ($q>0$) so that this force is repulsive \cite{moffat_svtg} and physically stable self-gravitating systems can exist \cite{moffat_galaxyrot1}. Additionally, $q$ is taken to be proportional to $m$, $q = \kappa m$ with $\kappa$ a positive proportionality constant \cite{moffat_eqmotion}, ensuring the validity of Einstein's Equivalence Principle.

The field equations associated to the MOG action in Eq. \eqref{eq:stvg-action} can be solved exactly assuming that:
\begin{enumerate}    
    \item the metric tensor is spherically symmetric;
    \item the scalar field $G$ can be treated as a constant on the scales of compact objects, $\partial_\nu G = 0$ \cite{moffat_blackhole1, Moffat_2021}. This means that the aforementioned parameter $\alpha$ can be regarded as a positive dimensionless constant, whose value depends on the mass of the gravitational source \citep{moffat_svtg}:
    \begin{equation}
        G = G_N(1+\alpha) = \textrm{const.}
        \label{eq:G}
    \end{equation}
    \item The proportionality constant $\kappa$ defining the fifth-force charge of massive particles is defined by:
    \begin{equation}
        \kappa = \sqrt{\alpha G_N}.
        \label{eq:kappa}
    \end{equation}
    \item The mass of the vector field, $\mu$, can be neglected on the scales of compact objects, as its effects are only evident on kpc scales \citep{moffat_galaxyrot1, moffat_galaxyrot2, moffat_clusters3};
\end{enumerate}
Under these assumptions (and by setting the speed of light in vacuum to $c= 1$), one obtains \cite{moffat_blackhole1} the following line element:
\begin{align}
    ds^2=&\frac{\Delta}{r^2}dt^2-\frac{r^2}{\Delta}dr^2-r^2d\Omega^2\,.
    \label{eq:stvg-sch-metric}
\end{align}
This Schwarzschild-like metric is the most general spherically symmetric static solution in MOG, and it provides with an exact description of the gravitational field around a point-like non-rotating source of mass $M$ (and hence a fifth-force charge $Q = \sqrt{\alpha G_N} M$). It differs from the classical one in GR (to which it reduces when $\alpha = 0$) by a different definition of the $\Delta$ function:
\begin{align}
    \Delta=r^2-2G_NMr+\alpha G_NM\bigl((1+\alpha)G_NM-2r\bigr).
    \label{eq:delta}
\end{align}
The solid angle element, on the other hand, has the usual expression $d\Omega^2 = d\theta^2+\sin^2\theta d\phi^2$. The vector field $\varphi^\alpha$ associated to the metric tensor in Eq. \eqref{eq:stvg-sch-metric} is given by \cite{lopez-romero}
\begin{equation}
    \varphi_\alpha = \left(-\frac{\sqrt{\alpha G_N}M}{r}, 0, 0, 0\right)
    \label{eq:vector_field},
\end{equation}
generating a repulsive force directed along the radial direction. As a consequence, the increased value of the gravitational constant $G$ increases the attractive effect of gravity on test particles, while the repulsive effect of the vector field counteracts this effect. As shown in \cite{rdm}, particles around a MOG BH experience an increased orbital precession, whose first-order expression explicitly depends on the parameter $\alpha$ and is given in Eq. \eqref{eq:stvg_precession}.

\section{The orbit of S2 in MOG}

\begin{figure}
    \begin{adjustwidth}{-\extralength}{0cm}
        \centering
        \includegraphics[width = \fulllength]{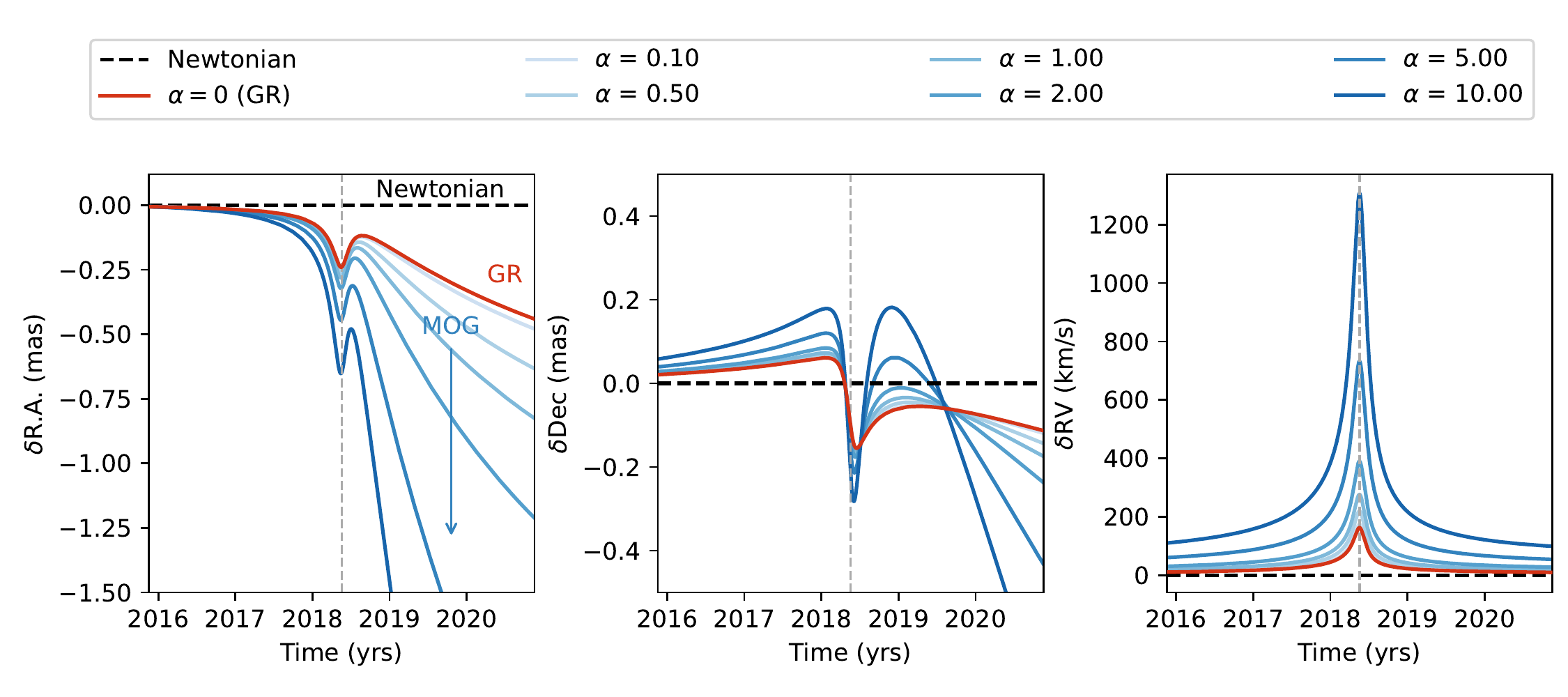}
    \end{adjustwidth}
    \caption{Numerically integrated sky-projected geodesic trajectories for the S2 star in pre- and post- pericentre (dashed vertical line) phase, for different values of the parameter $\alpha$. In particular, we report the deviation from the Newtonian case (dashed horizontal black line) of the GR orbit ($\alpha = 0$, red line) and for increasingly higher values of $\alpha$ (in different shades of blue) on the right ascension direction (\emph{left panel}), on the declination direction (\emph{central panel}) and for the radial velocity (\emph{right panel}).}
    \label{fig:deviation_from_GR}
\end{figure}

Upon integrating numerically the geodesic equations in Eq. \eqref{eq:geodesic-equations}, we obtain fully relativistic sky-projected orbits for the S2 star in MOG starting from its osculating Keplerian elements at the initial time\footnote{We refer to \cite{DellaMonica2021_a, DellaMonica2021_b} and the Supplementary Materials of \cite{deMartino2021} for a detailed description of our orbital model.}. These parameters are the semi-major axis of the orbit, $a$, the eccentricity $e$, the inclination $i$, the angle of the line of nodes $\Omega$, the angle from the ascending node to pericentre $\omega$, the orbital period $T$ and the time of the pericentre passage $t_p$. These uniquely assign the initial conditions of the star at a given time, that we set to be the time of passage at apocentre, given by $t_a = t_p-T/2$. Along with this parameters, one needs to fix the mass of the gravitational source, $M$, its distance from Earth, $R$, and a possible offset and drift (described by five additional parameters $x_0$, $y_0$, $v_{x,0}$, $v_{y,0}$ and $v_{z,0}$) of this object in the astrometric reference frame of the observer. From the integrated geodesic, the astrometric positions can be obtained via a geometric projection of the space-time coordinates, through the Thiele-Innes elements \cite{thieleinnes}, and modulating the observation times for the classical Rømer delay. The kinematic line-of-sight velocity of the star is converted into the spectroscopic observable, i.e. its redshift. In doing so we take into account both the special relativistic longitudinal and transverse Doppler effect and the gravitational time dilation, due to the combination of high-velocity and high-proximity at pericentre. Other effects, like the gravitational lensing or the Shapiro time delay give neglectable contributions \cite{deMartino2021, DellaMonica2021_b}, and we hence do not take them into account. In Figure \ref{fig:deviation_from_GR} we report how much the spectroscopic and the two astrometric observables deviate around pericentre from a Newtonian orbit of the S2 star, for different values of the parameter $\alpha$. As can be seen, measurements performed at and after pericentre of both the astrometric position of the star and its radial velocity carry a signature of the gravitational field produced in MOG.
    
\section{Data and methodology}

S2 is a B-type star in the nuclear star cluster of SgrA*, a compact radio source in the Galactic Centre (GC) of our Galaxy, identified with a supermassive black hole (SMBH) with mass $M\sim 4\times 10^6M_\odot$. Throughout its 16-year orbit , both special and general relativistic effects have been detected \cite{gravity2018, do, gravity} confirming predictions from GR, on one hand, and opening a new way to test gravity \cite{deMartino2021,DellaMonica2021_a,DellaMonica2021_b, jusufi2021}, on the other.

We exploit publicly available kinematic data for the S2 star to constrain the 15-dimensional parameter space of our orbital model in MOG, given by ($M$, $R$, $T$, $t_p$, $a$, $e$, $i$, $\Omega$, $\omega$, $x_0$, $y_0$, $v_{x,0}$, $v_{y,0}$, $v_{z, 0}$, $\alpha$). More specifically, we use the astrometric positions and radial velocities of S2 presented in \cite{gillessen} and the measurement of the relativistic orbital precession performed in 2020 by the Gravity Collaboration \cite{gravity}, through precise astrometric observations with the GRAVITY interferometer at VLT (which, however are not publicly available and we can only rely on the precession measurement itself). In particular, they measured the parameter $f_{\rm SP}$ in
    \begin{equation}
        \Delta\omega = \Delta\omega_{\rm GR}f_{\rm SP},
        \label{eq:gr_precession}
    \end{equation}
    where $\Delta\omega_{\rm GR}$ is given in Eq. \eqref{eq:gr_precession}, obtaining $f_{\rm sp}=1.10\pm 0.19$, thus favoring GR against Newtonian gravity at >5$\sigma$.

In order to fit our orbital model to such data we employ the Markov Chain Monte Carlo (MCMC) sampler in \texttt{emcee} \cite{emcee}, and we evaluate the integrated autocorrelation time of the chains to check the convergence of the algorithm. In particular, we perform two separate analyses:

\pgfkeys{/csteps/inner ysep=5pt}
\pgfkeys{/csteps/inner xsep=5pt}

\begin{enumerate}[itemsep=15pt, label=\Circled{\Alph*} :]
    \item We only use astrometric positions and radial velocities up to mid-2016 in \cite{gillessen}. Our dataset, thus, contains no information at all about the 2018 pericentre passage. In this case we use the following log-likelihood:
    \begin{equation}
        \log\mathcal{L} = \log\mathcal{N}(\textrm{R.A.}, \sigma_{\rm R.A.}) + \log\mathcal{N}(\textrm{Dec}, \sigma_{\rm Dec}) + \log\mathcal{N}(\textrm{RV}, \sigma_{\rm RV})
    \end{equation}
    by which we assume that all data points are uncorrelated with each other and that they are normally distributed within their experimental uncertainty, namely:
    \begin{equation}
        \log\mathcal{N}(\bm{x},\bm{\sigma}) = \sum_{i}\log\left[\frac{1}{\sigma_i\sqrt{2\pi}}\exp\left(\frac{(x_i-\mu_i)^2}{\sigma_i^2}\right)\right],
    \end{equation}  
    where $x_i$ is the $i$-th experimental data point, $\sigma_i$ its uncertainty and $\mu_i$ the corresponding prediction from our model.
    \item We use the same dataset used in case \Circled{A}, but adding as a single measurement the rate of orbital precession obtained in \cite{gravity}. Since the latter measurement was done using the same astrometric dataset that we use, plus data recorded at pericentre, we need to multiply all our uncertainties by $\sqrt{2}$ to avoid double counting data points. This yields
    \begin{adjustwidth}{-\extralength}{0cm}
        \begin{equation}
            \log\mathcal{L} = \log\mathcal{N}(\textrm{R.A.}, \sqrt{2}\sigma_{\rm R.A.}) + \log\mathcal{N}(\textrm{Dec}, \sqrt{2}\sigma_{\rm Dec}) + \log\mathcal{N}(\textrm{RV}, \sqrt{2}\sigma_{\rm RV}) +\log\mathcal{N}(f_{\rm SP}, \sqrt{2}\sigma_{\rm SP}).
        \end{equation}
    \end{adjustwidth}
\end{enumerate}
\vspace{0.5cm}
In both cases we use uniform flat priors for our parameters\footnote{Except for the five reference frame parameters, $x_0$, $y_0$, $v_{x,0}$, $v_{y,0}$ and $v_{z, 0}$, for which we use gaussian priors from the independent measurements by \cite{Plewa}, and for the parameters $T$ and $t_p$ we use large (FWHM = 10 times the experimental uncertainty) gaussian priors centered on their best fitting values from \cite{gravity}.} centered on their best-fitting value by \cite{gillessen} and with an amplitude given by 10 times their experimental uncertainty, and we set heuristically $\alpha \in [0, 2]$ as uniform interval for the MOG parameter.

\section{Results}

\begin{figure}
    \begin{adjustwidth}{-\extralength}{0cm}
        \centering
        \includegraphics[width = \fulllength]{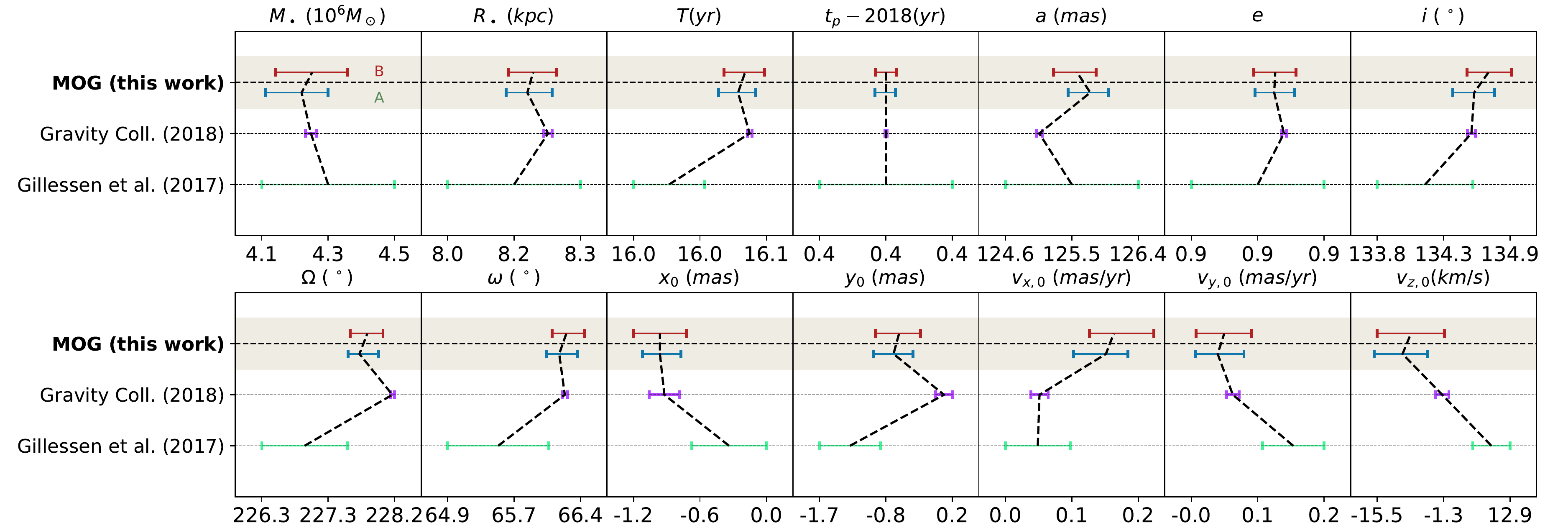}
    \end{adjustwidth}
    \caption{The best fitting values and $1\sigma$ confidence interval for the orbital parameters of the S2 star in our analyses \Circled{A} (blue bars) and \Circled{B} (red bars), compared with the best-fitting values from previous works in \cite{gravity} (in which 1-PPN model is fitted to the data) and \cite{gillessen} (using a Keplerian model to describe the orbit of S2).}
    \label{fig:fit_results}
\end{figure}

\begin{figure}
    \centering
    \includegraphics[width = 0.6\textwidth]{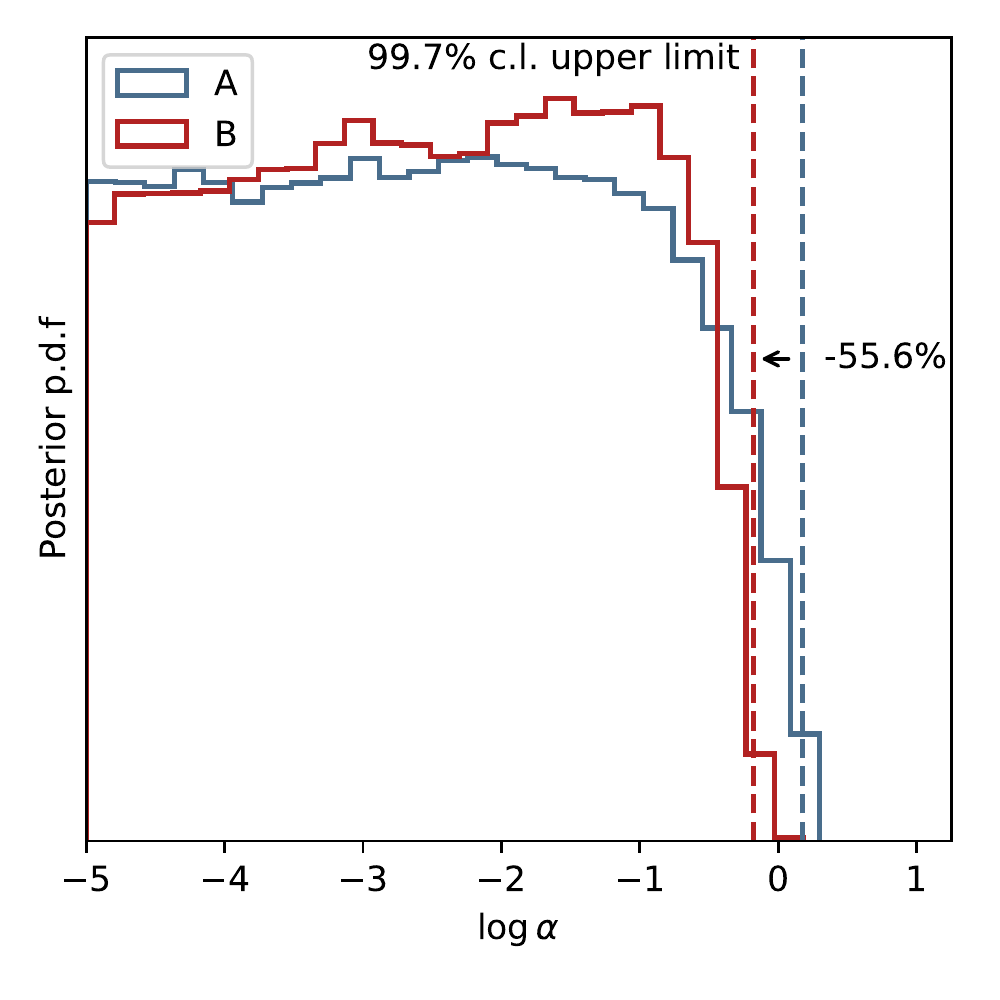}
    \caption{The normalized posterior probability distribution of the parameter $\alpha$ in logarithmic scale for the two analysis \Circled{A} (in blue) and \Circled{B} (in red). The 99.7\% c.l. level upper limit of the parameter is reported as a dashed vertical line in the two cases. The analysis \Circled{B} provides with a more peaked distribution for $\alpha$ around 0, with the upper limit going down by $55.6\%$ between the two analyses.}
    \label{fig:alpha_results}
\end{figure}

In Figure \ref{fig:fit_results} we report the $1\sigma$ confidence intervals for the orbital parameters in our analyses \Circled{A} and \Circled{B}, compared with the corresponding $1\sigma$ intervals from \cite{gillessen} (who fitted Keplerian orbits to the data) and \cite{gravity} (in which a first-order Post-Newtonian orbital model is used). The parameters from our analyses are compatible within their errors with the results from the previous studies. Finally, in Figure \ref{fig:alpha_results} we report in logarithmic scale the normalized posterior distributions for the parameter $\alpha$ from the two analysis \Circled{A} (in blue) and \Circled{B} (in red) along with their 99.7\% confidence level (c.l.) upper limit. Our results provide with the first constraint on the MOG theory at the GC, yielding:
\begin{align}
    \textrm{\Circled{A} :}&\qquad \alpha \lesssim 1.499 \qquad\textrm{w/o precession}\\
    \textrm{\Circled{B} :}&\qquad \alpha \lesssim 0.662 \qquad\textrm{w/ precession}
\end{align}
While both analyses are compatible with GR, the additional information carried by the single orbital precession data point at pericentre results in a more peaked distribution for $\alpha$ in case \Circled{B}, whose upper limit decrease by $55.6\%$ with respect to analysis \Circled{A}.

\section{Conclusions}

Here, we have summarized our results in \cite{DellaMonica2021_a} providing with the first constraint on the extra parameter, $\alpha$, of MOG at the GC, obtained by studying the fully-relativistic motion of the S2 star around the SMBH SgrA*. In particular, we have solved numerically the geodesic equations for a test particle around a static BH in MOG, described by the metric element in Eq. \eqref{eq:stvg-sch-metric}, particularizing the kinematic properties of the test particle for the orbital parameters of the S2 star \cite{gillessen,gravity}. Then, we have explored the 15-dimensional parameter space of our model by means of a MCMC algorithm, which allowed us to study the posterior distributions of the parameters, upon comparison with publicly available kinematic data for S2 and the measurement of its rate of orbital precession. In particular, we have performed two separate analyses, in which we have either excluded - \Circled{A} - or included - \Circled{B} - the information of the pericentre advance. We have demonstrated that the additional information carried by the relativistic orbital precession is able to bring down the 99.7$\%$ c.l. upper limit on the parameter $\alpha$ from $\alpha \lesssim 1.499$ in analysis \Circled{A} to $\alpha \lesssim 0.662$ in analysis \Circled{B}. A significant reduction of $\sim55.6\%$.

\vspace{6pt} 

\funding{RDM acknowledges support from Consejeria de Educación de la Junta de Castilla y León and from the Fondo Social Europeo.
IDM acknowledges support from Ayuda  IJCI2018-036198-I  funded by  MCIN/AEI/  10.13039/501100011033  and  FSE  “El FSE  invierte  en  tu  futuro”  o  “Financiado  por  la  Unión  Europea   “NextGenerationEU”/PRTR. 
IDM is also supported by the project PGC2018-096038-B-I00  funded by the Spanish "Ministerio de Ciencia e Innovación" and FEDER “A way of making Europe", and by the project SA096P20 Junta de Castilla y León. M.D.L. acknowledges INFN Sez. di Napoli (Iniziativa Specifica TEONGRAV). 
}

 \dataavailability{Data can be found in \cite{gillessen}.} 


 \conflictsofinterest{The authors declare no conflict of interest.} 

\begin{adjustwidth}{-\extralength}{0cm}

\reftitle{References}



\bibliography{biblio}

\begin{thebibliography}{999}

\bibitem[{Moffat}(2006)]{moffat_svtg}
{Moffat}, J.W.
\newblock {Scalar tensor vector gravity theory}.
\newblock {\em \jcap} {\bf 2006}, {\em 2006},~004,
  \href{http://xxx.lanl.gov/abs/gr-qc/0506021}{{\normalfont
  [arXiv:gr-qc/gr-qc/0506021]}}.
\newblock
  doi:{\changeurlcolor{black}\href{https://doi.org/10.1088/1475-7516/2006/03/004}{\detokenize{10.1088/1475-7516/2006/03/004}}}.

\bibitem[{Brownstein} and {Moffat}(2006)]{Brownstein2006}
{Brownstein}, J.R.; {Moffat}, J.W.
\newblock {Galaxy Rotation Curves without Nonbaryonic Dark Matter}.
\newblock {\em \apj} {\bf 2006}, {\em 636},~721--741,
  \href{http://xxx.lanl.gov/abs/astro-ph/0506370}{{\normalfont
  [arXiv:astro-ph/astro-ph/0506370]}}.
\newblock
  doi:{\changeurlcolor{black}\href{https://doi.org/10.1086/498208}{\detokenize{10.1086/498208}}}.

\bibitem[{Moffat} and {Rahvar}(2013)]{moffat_galaxyrot1}
{Moffat}, J.W.; {Rahvar}, S.
\newblock {The MOG weak field approximation and observational test of galaxy
  rotation curves}.
\newblock {\em \mnras} {\bf 2013}, {\em 436},~1439--1451,
  \href{http://xxx.lanl.gov/abs/1306.6383}{{\normalfont
  [arXiv:astro-ph.GA/1306.6383]}}.
\newblock
  doi:{\changeurlcolor{black}\href{https://doi.org/10.1093/mnras/stt1670}{\detokenize{10.1093/mnras/stt1670}}}.

\bibitem[{Moffat} and {Toth}(2015)]{moffat_galaxyrot2}
{Moffat}, J.W.; {Toth}, V.T.
\newblock {Rotational velocity curves in the Milky Way as a test of modified
  gravity}.
\newblock {\em \prd} {\bf 2015}, {\em 91},~043004,
  \href{http://xxx.lanl.gov/abs/1411.6701}{{\normalfont
  [arXiv:astro-ph.GA/1411.6701]}}.
\newblock
  doi:{\changeurlcolor{black}\href{https://doi.org/10.1103/PhysRevD.91.043004}{\detokenize{10.1103/PhysRevD.91.043004}}}.

\bibitem[{De Martino} and {De Laurentis}(2017)]{DeMartino2017}
{De Martino}, I.; {De Laurentis}, M.
\newblock {On the universality of MOG weak field approximation at galaxy
  cluster scale}.
\newblock {\em Physics Letters B} {\bf 2017}, {\em 770},~440--444,
  \href{http://xxx.lanl.gov/abs/1705.02366}{{\normalfont
  [arXiv:astro-ph.CO/1705.02366]}}.
\newblock
  doi:{\changeurlcolor{black}\href{https://doi.org/10.1016/j.physletb.2017.05.012}{\detokenize{10.1016/j.physletb.2017.05.012}}}.

\bibitem[{de Martino}(2020)]{deMartino2020}
{de Martino}, I.
\newblock {Giant low-surface-brightness dwarf galaxy as a test bench for
  MOdified Gravity}.
\newblock {\em \mnras} {\bf 2020}, {\em 493},~2373--2376,
  \href{http://xxx.lanl.gov/abs/2002.05161}{{\normalfont
  [arXiv:gr-qc/2002.05161]}}.
\newblock
  doi:{\changeurlcolor{black}\href{https://doi.org/10.1093/mnras/staa460}{\detokenize{10.1093/mnras/staa460}}}.

\bibitem[{de Martino} \em{et~al.}(2020){de Martino}, {Chakrabarty}, {Cesare},
  {Gallo}, {Ostorero}, and {Diaferio}]{deMartinoChakrabarty2020}
{de Martino}, I.; {Chakrabarty}, S.S.; {Cesare}, V.; {Gallo}, A.; {Ostorero},
  L.; {Diaferio}, A.
\newblock {Dark Matters on the Scale of Galaxies}.
\newblock {\em Universe} {\bf 2020}, {\em 6},~107,
  \href{http://xxx.lanl.gov/abs/2007.15539}{{\normalfont
  [arXiv:astro-ph.CO/2007.15539]}}.
\newblock
  doi:{\changeurlcolor{black}\href{https://doi.org/10.3390/universe6080107}{\detokenize{10.3390/universe6080107}}}.

\bibitem[{Della Monica} \em{et~al.}(2021){Della Monica}, {de Martino}, and {de
  Laurentis}]{DellaMonica2021_a}
{Della Monica}, R.; {de Martino}, I.; {de Laurentis}, M.
\newblock {Orbital precession of the S2 star in Scalar-Tensor-Vector-Gravity}.
\newblock {\em \mnras} {\bf 2021}.
\newblock
  doi:{\changeurlcolor{black}\href{https://doi.org/10.1093/mnras/stab3727}{\detokenize{10.1093/mnras/stab3727}}}.

\bibitem[Della~Monica \em{et~al.}(2021)Della~Monica, De~Laurentis, and
  Younsi]{rdm}
Della~Monica, R.; De~Laurentis, M.; Younsi, Z.
\newblock STVG-shadow and particle motion.
\newblock {\em In preparation} {\bf 2021}.

\bibitem[{Gillessen} \em{et~al.}(2017){Gillessen}, {Plewa}, {Eisenhauer},
  {Sari}, {Waisberg}, {Habibi}, {Pfuhl}, {George}, {Dexter}, {von Fellenberg},
  {Ott}, and {Genzel}]{gillessen}
{Gillessen}, S.; {Plewa}, P.M.; {Eisenhauer}, F.; {Sari}, R.; {Waisberg}, I.;
  {Habibi}, M.; {Pfuhl}, O.; {George}, E.; {Dexter}, J.; {von Fellenberg}, S.;
  et~al.
\newblock {An Update on Monitoring Stellar Orbits in the Galactic Center}.
\newblock {\em \apj} {\bf 2017}, {\em 837},~30,
  \href{http://xxx.lanl.gov/abs/1611.09144}{{\normalfont
  [arXiv:astro-ph.GA/1611.09144]}}.
\newblock
  doi:{\changeurlcolor{black}\href{https://doi.org/10.3847/1538-4357/aa5c41}{\detokenize{10.3847/1538-4357/aa5c41}}}.

\bibitem[{Gravity Collaboration} \em{et~al.}(2020){Gravity Collaboration},
  {Abuter}, {Amorim}, {Baub{\"o}ck}, {Berger}, {Bonnet}, {Brandner}, {Cardoso},
  {Cl{\'e}net}, {de Zeeuw}, {Dexter}, {Eckart}, {Eisenhauer}, {F{\"o}rster
  Schreiber}, {Garcia}, {Gao}, {Gendron}, {Genzel}, {Gillessen}, {Habibi},
  {Haubois}, {Henning}, {Hippler}, {Horrobin}, {Jim{\'e}nez-Rosales}, {Jochum},
  {Jocou}, {Kaufer}, {Kervella}, {Lacour}, {Lapeyr{\`e}re}, {Le Bouquin},
  {L{\'e}na}, {Nowak}, {Ott}, {Paumard}, {Perraut}, {Perrin}, {Pfuhl},
  {Rodr{\'\i}guez-Coira}, {Shangguan}, {Scheithauer}, {Stadler}, {Straub},
  {Straubmeier}, {Sturm}, {Tacconi}, {Vincent}, {von Fellenberg}, {Waisberg},
  {Widmann}, {Wieprecht}, {Wiezorrek}, {Woillez}, {Yazici}, and
  {Zins}]{gravity}
{Gravity Collaboration}.; {Abuter}, R.; {Amorim}, A.; {Baub{\"o}ck}, M.;
  {Berger}, J.P.; {Bonnet}, H.; {Brandner}, W.; {Cardoso}, V.; {Cl{\'e}net},
  Y.; {de Zeeuw}, P.T.;  et~al.
\newblock {Detection of the Schwarzschild precession in the orbit of the star
  S2 near the Galactic centre massive black hole}.
\newblock {\em \aap} {\bf 2020}, {\em 636},~L5,
  \href{http://xxx.lanl.gov/abs/2004.07187}{{\normalfont
  [arXiv:astro-ph.GA/2004.07187]}}.
\newblock
  doi:{\changeurlcolor{black}\href{https://doi.org/10.1051/0004-6361/202037813}{\detokenize{10.1051/0004-6361/202037813}}}.

\bibitem[{Moffat} and {Toth}(2009)]{moffat_eqmotion}
{Moffat}, J.W.; {Toth}, V.T.
\newblock {Fundamental parameter-free solutions in modified gravity}.
\newblock {\em Classical and Quantum Gravity} {\bf 2009}, {\em 26},~085002,
  \href{http://xxx.lanl.gov/abs/0712.1796}{{\normalfont
  [arXiv:gr-qc/0712.1796]}}.
\newblock
  doi:{\changeurlcolor{black}\href{https://doi.org/10.1088/0264-9381/26/8/085002}{\detokenize{10.1088/0264-9381/26/8/085002}}}.

\bibitem[{Moffat}(2015)]{moffat_blackhole1}
{Moffat}, J.W.
\newblock {Black holes in modified gravity (MOG)}.
\newblock {\em European Physical Journal C} {\bf 2015}, {\em 75},~175,
  \href{http://xxx.lanl.gov/abs/1412.5424}{{\normalfont
  [arXiv:gr-qc/1412.5424]}}.
\newblock
  doi:{\changeurlcolor{black}\href{https://doi.org/10.1140/epjc/s10052-015-3405-x}{\detokenize{10.1140/epjc/s10052-015-3405-x}}}.

\bibitem[{Moffat}(2021)]{Moffat_2021}
{Moffat}, J.W.
\newblock {Modified gravity (MOG), cosmology and black holes}.
\newblock {\em \jcap} {\bf 2021}, {\em 2021},~017,
  \href{http://xxx.lanl.gov/abs/2006.12550}{{\normalfont
  [arXiv:gr-qc/2006.12550]}}.
\newblock
  doi:{\changeurlcolor{black}\href{https://doi.org/10.1088/1475-7516/2021/02/017}{\detokenize{10.1088/1475-7516/2021/02/017}}}.

\bibitem[{Brownstein} and {Moffat}(2007)]{moffat_clusters3}
{Brownstein}, J.R.; {Moffat}, J.W.
\newblock {The Bullet Cluster 1E0657-558 evidence shows modified gravity in the
  absence of dark matter}.
\newblock {\em \mnras} {\bf 2007}, {\em 382},~29--47,
  \href{http://xxx.lanl.gov/abs/astro-ph/0702146}{{\normalfont
  [arXiv:astro-ph/astro-ph/0702146]}}.
\newblock
  doi:{\changeurlcolor{black}\href{https://doi.org/10.1111/j.1365-2966.2007.12275.x}{\detokenize{10.1111/j.1365-2966.2007.12275.x}}}.

\bibitem[{Lopez Armengol} and {Romero}(2017)]{lopez-romero}
{Lopez Armengol}, F.G.; {Romero}, G.E.
\newblock {Neutron stars in Scalar-Tensor-Vector Gravity}.
\newblock {\em General Relativity and Gravitation} {\bf 2017}, {\em 49},~27,
  \href{http://xxx.lanl.gov/abs/1611.05721}{{\normalfont
  [arXiv:gr-qc/1611.05721]}}.
\newblock
  doi:{\changeurlcolor{black}\href{https://doi.org/10.1007/s10714-017-2184-0}{\detokenize{10.1007/s10714-017-2184-0}}}.

\bibitem[{Della Monica} and {de Martino}(2021)]{DellaMonica2021_b}
{Della Monica}, R.; {de Martino}, I.
\newblock {Unveiling the nature of SgrA* with the geodesic motion of S-stars}.
\newblock {\em arXiv e-prints} {\bf 2021}, p. arXiv:2112.01888,
  \href{http://xxx.lanl.gov/abs/2112.01888}{{\normalfont
  [arXiv:astro-ph.GA/2112.01888]}}.

\bibitem[{De Martino} \em{et~al.}(2021){De Martino}, {della Monica}, and {De
  Laurentis}]{deMartino2021}
{De Martino}, I.; {della Monica}, R.; {De Laurentis}, M.
\newblock {f (R ) gravity after the detection of the orbital precession of the
  S2 star around the Galactic Center massive black hole}.
\newblock {\em \prd} {\bf 2021}, {\em 104},~L101502,
  \href{http://xxx.lanl.gov/abs/2106.06821}{{\normalfont
  [arXiv:gr-qc/2106.06821]}}.
\newblock
  doi:{\changeurlcolor{black}\href{https://doi.org/10.1103/PhysRevD.104.L101502}{\detokenize{10.1103/PhysRevD.104.L101502}}}.

\bibitem[{Taff} and {Szebehely}(1986)]{thieleinnes}
{Taff}, L.G.; {Szebehely}, V.
\newblock {Book-Review - Celestial Mechanics - a Computational Guide for the
  Practitioner}.
\newblock {\em \nat} {\bf 1986}, {\em 319},~630.

\bibitem[{Gravity Collaboration} \em{et~al.}(2018){Gravity Collaboration},
  {Abuter}, {Amorim}, {Anugu}, {Baub{\"o}ck}, {Benisty}, {Berger}, {Blind},
  {Bonnet}, {Brandner}, {Buron}, {Collin}, {Chapron}, {Cl{\'e}net}, {Coud{\'e}
  Du Foresto}, {de Zeeuw}, {Deen}, {Delplancke-Str{\"o}bele}, {Dembet},
  {Dexter}, {Duvert}, {Eckart}, {Eisenhauer}, {Finger}, {F{\"o}rster
  Schreiber}, {F{\'e}dou}, {Garcia}, {Garcia Lopez}, {Gao}, {Gendron},
  {Genzel}, {Gillessen}, {Gordo}, {Habibi}, {Haubois}, {Haug}, {Hau{\ss}mann},
  {Henning}, {Hippler}, {Horrobin}, {Hubert}, {Hubin}, {Jimenez Rosales},
  {Jochum}, {Jocou}, {Kaufer}, {Kellner}, {Kendrew}, {Kervella}, {Kok},
  {Kulas}, {Lacour}, {Lapeyr{\`e}re}, {Lazareff}, {Le Bouquin}, {L{\'e}na},
  {Lippa}, {Lenzen}, {M{\'e}rand}, {M{\"u}ler}, {Neumann}, {Ott}, {Palanca},
  {Paumard}, {Pasquini}, {Perraut}, {Perrin}, {Pfuhl}, {Plewa}, {Rabien},
  {Ram{\'\i}rez}, {Ramos}, {Rau}, {Rodr{\'\i}guez-Coira}, {Rohloff}, {Rousset},
  {Sanchez-Bermudez}, {Scheithauer}, {Sch{\"o}ller}, {Schuler}, {Spyromilio},
  {Straub}, {Straubmeier}, {Sturm}, {Tacconi}, {Tristram}, {Vincent}, {von
  Fellenberg}, {Wank}, {Waisberg}, {Widmann}, {Wieprecht}, {Wiest},
  {Wiezorrek}, {Woillez}, {Yazici}, {Ziegler}, and {Zins}]{gravity2018}
{Gravity Collaboration}.; {Abuter}, R.; {Amorim}, A.; {Anugu}, N.;
  {Baub{\"o}ck}, M.; {Benisty}, M.; {Berger}, J.P.; {Blind}, N.; {Bonnet}, H.;
  {Brandner}, W.;  et~al.
\newblock {Detection of the gravitational redshift in the orbit of the star S2
  near the Galactic centre massive black hole}.
\newblock {\em \aap} {\bf 2018}, {\em 615},~L15,
  \href{http://xxx.lanl.gov/abs/1807.09409}{{\normalfont
  [arXiv:astro-ph.GA/1807.09409]}}.
\newblock
  doi:{\changeurlcolor{black}\href{https://doi.org/10.1051/0004-6361/201833718}{\detokenize{10.1051/0004-6361/201833718}}}.

\bibitem[{Do} \em{et~al.}(2019){Do}, {Hees}, {Ghez}, {Martinez}, {Chu}, {Jia},
  {Sakai}, {Lu}, {Gautam}, {O'Neil}, {Becklin}, {Morris}, {Matthews},
  {Nishiyama}, {Campbell}, {Chappell}, {Chen}, {Ciurlo}, {Dehghanfar},
  {Gallego-Cano}, {Kerzendorf}, {Lyke}, {Naoz}, {Saida}, {Sch{\"o}del},
  {Takahashi}, {Takamori}, {Witzel}, and {Wizinowich}]{do}
{Do}, T.; {Hees}, A.; {Ghez}, A.; {Martinez}, G.D.; {Chu}, D.S.; {Jia}, S.;
  {Sakai}, S.; {Lu}, J.R.; {Gautam}, A.K.; {O'Neil}, K.K.;  et~al.
\newblock {Relativistic redshift of the star S0-2 orbiting the Galactic Center
  supermassive black hole}.
\newblock {\em Science} {\bf 2019}, {\em 365},~664--668,
  \href{http://xxx.lanl.gov/abs/1907.10731}{{\normalfont
  [arXiv:astro-ph.GA/1907.10731]}}.
\newblock
  doi:{\changeurlcolor{black}\href{https://doi.org/10.1126/science.aav8137}{\detokenize{10.1126/science.aav8137}}}.

\bibitem[{Jusufi} \em{et~al.}(2021){Jusufi}, {Azreg-A{\"\i}nou}, {Jamil}, and
  {Saridakis}]{jusufi2021}
{Jusufi}, K.; {Azreg-A{\"\i}nou}, M.; {Jamil}, M.; {Saridakis}, E.N.
\newblock {Constraints on Barrow entropy from M87* and S2 star observations}.
\newblock {\em arXiv e-prints} {\bf 2021}, p. arXiv:2110.07258,
  \href{http://xxx.lanl.gov/abs/2110.07258}{{\normalfont
  [arXiv:gr-qc/2110.07258]}}.

\bibitem[{Foreman-Mackey} \em{et~al.}(2013){Foreman-Mackey}, {Hogg}, {Lang},
  and {Goodman}]{emcee}
{Foreman-Mackey}, D.; {Hogg}, D.W.; {Lang}, D.; {Goodman}, J.
\newblock {emcee: The MCMC Hammer}.
\newblock {\em Publications of the Astronomical Society of the Pacific} {\bf
  2013}, {\em 125},~306,  \href{http://xxx.lanl.gov/abs/1202.3665}{{\normalfont
  [arXiv:astro-ph.IM/1202.3665]}}.
\newblock
  doi:{\changeurlcolor{black}\href{https://doi.org/10.1086/670067}{\detokenize{10.1086/670067}}}.

\bibitem[{Plewa} \em{et~al.}(2015){Plewa}, {Gillessen}, {Eisenhauer}, {Ott},
  {Pfuhl}, {George}, {Dexter}, {Habibi}, {Genzel}, {Reid}, and {Menten}]{Plewa}
{Plewa}, P.M.; {Gillessen}, S.; {Eisenhauer}, F.; {Ott}, T.; {Pfuhl}, O.;
  {George}, E.; {Dexter}, J.; {Habibi}, M.; {Genzel}, R.; {Reid}, M.J.;  et~al.
\newblock {Pinpointing the near-infrared location of Sgr A* by correcting
  optical distortion in the NACO imager}.
\newblock {\em \mnras} {\bf 2015}, {\em 453},~3234--3244,
  \href{http://xxx.lanl.gov/abs/1509.01941}{{\normalfont
  [arXiv:astro-ph.GA/1509.01941]}}.
\newblock
  doi:{\changeurlcolor{black}\href{https://doi.org/10.1093/mnras/stv1910}{\detokenize{10.1093/mnras/stv1910}}}.

\end{thebibliography}

\end{adjustwidth}
\end{document}